\documentclass[aps,prd,onecolumn,groupedaddress,nofootinbib,showpacs,showkeys,amssymb,11pt]{revtex4}
\usepackage[a4paper,left=25mm,right=25mm,top=25mm,bottom=25mm]{geometry}
\usepackage{epsfig} 
\usepackage{amsmath} 
\usepackage{amsfonts}
\usepackage{amssymb} 
\usepackage{graphicx} 
\usepackage{colordvi}

\def\be{\begin{equation}}
\def\ee{\end{equation}}
\def\bea{\begin{eqnarray}}
\def\eea{\end{eqnarray}}

\renewcommand{\epsilon}{\varepsilon}

\def\beqa{\begin{eqnarray}}
\def\eeqa{\end{eqnarray}}
\def\beq{\begin{equation}}
\def\eeq{\end{equation}}

\renewcommand{\epsilon}{\varepsilon}

\def\({\left(}    \def\){\right)}

\def\data{\the\day-\the\month-\the\year}

\def\frac#1#2{{#1 \over #2}}

\def\phi{\varphi}

\def\~{\approx}

\usepackage{latexsym,float}
\usepackage{epsfig,graphics}
\usepackage{epstopdf}
\usepackage{amssymb}
\usepackage{amsmath}
\usepackage{verbatim}
\usepackage{multirow}
\usepackage{color}
\usepackage{tikz}
\usepackage[utf8]{inputenc}
\usepackage{commath}
\usepackage{braket}
\usepackage{amssymb}
\usepackage{lineno}
\usepackage{graphicx}

\def\gtwid{\mathrel{\raise.3ex\hbox{$>$\kern-.75em\lower1ex\hbox{$\sim$}}}}
\def\ltwid{\mathrel{\raise.3ex\hbox{$<$\kern-.75em\lower1ex\hbox{$\sim$}}}}
\def\square{\kern1pt\vbox{\hrule height 1.2pt\hbox{\vrule width 1.2pt\hskip 3pt
			\vbox{\vskip 6pt}\hskip 3pt\vrule width 0.6pt}\hrule height 0.6pt}\kern1pt}
\begin{document}
\title{ Holographic Entanglement Entropy, Complexity, Fidelity Susceptibility and Hierarchical UV/IR Mixing Problem in $AdS_2/\mbox{open strings}$  }

 \author{Kazuharu Bamba}\email{bamba@sss.fukushima-u.ac.jp}
\affiliation{{
		Division of Human Support System, Faculty of Symbiotic Systems Science, Fukushima University,\\ Fukushima 960-1296, Japan}}      
\author{Davood Momeni} \email{davood@squ.edu.om}
\affiliation{{Department of Physics, College of Science, Sultan Qaboos University,
		\\P.O. Box 36, P.C. 123,  Muscat, Sultanate of Oman}}
\author{Mudhahir Al Ajmi} \email{mudhahir@squ.edu.om}
\affiliation{{Department of Physics, College of Science, Sultan Qaboos University,
		\\P.O. Box 36, P.C. 123,  Muscat, Sultanate of Oman}}



\begin{abstract}
In this paper, we will compute the holographic complexity (dual to a volume in AdS), holographic fidelity susceptibility and the holographic entanglement entropy (dual to an area in AdS) in a two-dimensional version of $AdS$ which is dual to open strings. We will explicitly demonstrate that these quantities are well defined, and then argue that a relation for fidelity susceptibility and time should hold in general due to the  $AdS_2$  version of the classical Kepler's  principle. We will demonstrate that it holds for $AdS_2$ solution as well as conformal copies metrics in bulk theory of a prescribed dual conformal invariant quantum mechanics which have been obtained in open string theory. We will also show that hierarchical UV/IR mixing exists in boundary string theory through the holographic bulk picture.
\end{abstract}
\date{\today}

\maketitle
\tableofcontents 
\section{Introduction}
\label{sec:intro}

Varied studies done in wide areas of  physics have proved that  the fundamental  laws of physics can be  reformulated   in terms of 
 relevant information theory quantities \cite{info, info2}.  
Entropy quantifies the amount of information bits that are  lost in a certain (un)physical process, and thus, it 
is supposed to be one of the most important physical quantities related to any such information theoretical process. 
The entropy has been used to realize several physical phenomena from condensed matter physics (like phase transitions, critical phenomena) to gravitational physics, where entropy looks just like the area of certain surfaces, called horizons. Also, it is believed that the geometry  of spacetime  can be 
underestood as an emergent object, which emerges  due to 
some types of  information theoretical process. A simple reason to believe it is that 
,  in the Jacobson formalism where gravitational field equations are related to thermodynamical quantities, it is always possible to derive the Einstein equations  from thermodynamics of horizon by presuming a certain scaling form 
for  the  entropy \cite{z12j, jz12}. We know that the  maximum entropy of a certain region of space scales with the horizon's 
area, although this observation has been  acquired holding the physics  of black holes.  This naive realtion between area (boundary) and entropy (a quantity relates to the quantum states inside a system) makes the idea of
holographic principle \cite{1, 2}, and the AdS/CFT correspondence, one of the most significant dualities between two regimes (strong/weak) of several physical systems  \cite{M:1997}. 

The AdS/CFT correspondence makes it possible to describe   quantum 
entanglement in complex systems in the form  of the holographic
entanglement entropy (HEE) \cite{6,RT,6a}.

  The HEE of a given quantum field theory in $d+1$ dimensions (even non relativistic on) is holographically calculated in terms of  the area of a minimal surface defined in the geometry  of an asymptotically $AdS_d$  dual geometry. Let us consider  
HEE for a given subsystem $A$ with its complement $A'$. The Ryu-Takayanagi (RT)  expression for the holographic entanglement entropy is
\begin{equation} \label{HEE}
S_{A}=\frac{\mathcal{A}(\gamma _{A})}{4G_{d+1}}
\end{equation}
where  $G$  is the gravitational constant in $d$ dimensions,  $\gamma_{A}$ is the $(d-1)$-minimal surface in the  $AdS_d$ geometry. We assumed that  the boundary of this surface named as  $\mathcal{A}(\gamma _{A})$ is the dsame as the boundary of the quantum entangled system $\partial A$. 
Because of non renormalizability of Einstein gravity as well as existence of cutoffs, in RT scheme to compute HEE, there are UV divergence terms  like $\epsilon^{-n},n\geq1, \ln\epsilon,..,$ etc. Consequently we are required  to find
a regularization strategy to improve (remove) these divergences. Inspired from quantum field theory, we consider a  deformed geometry $D$ and we define the area as follows,
\begin{eqnarray}
 \mathcal{A}(\gamma_A) = \mathcal{A}_{D}(\gamma_A) - \mathcal{A}_{AdS}(\gamma_A), 
\end{eqnarray}
where $ \mathcal{A}_{D}(\gamma_A) $ is  defined in deformed geometry (for example excited states), and $ \mathcal{A}_{AdS}(\gamma_A)$
is defined in the background $AdS$ spacetime (ground state). 
We hope that if one defines 
the holographic entanglement entropy for a deformed geometry by subtracting the 
contribution coming from the background $AdS$ spacetime, we are only left with a finite part. 
We will use this scheme of renormalization through this paper.
\par 
As mentioned, the entropy measures the amount of the information which we lost during a physical process. It is very common to define the complexity as a quantity which quantifies the difficulty to obtain the information of a system. 
The complexity  has been introduced  to investigate different physical systems from gravitational physics to condensed matter physics, and even 
quantum information theory. This lost information never can be retracted using any possible physical process  \cite{hawk}. 
Because complexity has only been recently  introduced to investigate miscellaneous physical systems, there are different schemes to define the 
complexity for a CFT. However, recently 
inspired  by  RT proposal for holographic entanglement entropy,  
 holographic complexity (HC) has been holographically conjectured as a certsain types of volumes in the dual  anti-de Sitter (AdS) background \cite{Susskind:2014rva1} -\cite{Stanford:2014jda}.
Moreover, it is adequate to specify  a subsystem $A$ with its complement, and define this volume as  $V = V(\gamma_A)$, 
i.e., the volume enclosed by the same  minimal surface which was prposed  to estimate the HEE \cite{Alishahiha:2015rta} is given as follows (called holographic complexity or HC),
\begin{equation}\label{HC}
\mathcal{C}_A= \frac{V(\gamma_A) }{8\pi R G_{d+1}},
\end{equation}
here $R$ and $V(\gamma_A)$ are the radius of the curvature and the volume in the $AdS_d$ bulk geometry. This volume  contains UV divergences, and  so we need an appropriate
regularization scheme  for it. In analogous to the HEE,  we define the regularized volume as 
\begin{eqnarray}
\Delta\mathcal{V}(\gamma_A) = V_{D}(\gamma_A) - V_{AdS}(\gamma_A).
\end{eqnarray}
Here $ \Delta\mathcal{V}_{D}(\gamma_A) $ denotes the volume  in deformed geometry,  and $V_{AdS}(\gamma_A)$
is the volume in the background $AdS$ spacetime.  This again improves the divergences  and one is again  left with a finite part. 
Several examples for HC have been studied in literature  \cite{Momeni:2016ekm}-\cite{Momeni:2016qfv}. 

A  type of duality between dilaton gravity (gravity in 2 dimensions) on $AdS_2$ and open strings was discovred in \cite{Cadoni:2000gm} and it was clearly shown how $AdS_2$ is equivalent to conformal quantum mechanics (CQM) as a non relativistic limit of $CFT_1$ (see \cite{Cadoni:2000gm} and other papers of these authors). This lower dimensional version of $AdS_{d+1}/CFT_d$  conjecture argued that gravity on $AdS_2$ is holographically dual to a one-dimensional conformal field theory on the  boundary of $AdS_2$. One reason to believe that such duality exists, is due to the fact that classical two-dimensional (dilaton) gauge theory for gravity has trivial  conformal symmetry. It can be possible to reformulate this gauge theory  as a nonlinear sigma-model \cite{M. Cavagli`a} and it was proved that when we take trhe classical limit of this toy model,  we were observed the  conformal symmetry. This is a reason to think about  gravity on $AdS_2$ as a natural  dual to a one-dimensional CFT  Entanglement entropy, Holographic Complexity and Fidelity Susceptibility in open strings. This duality is called as
$AdS_2/\mbox{open string}$ or $AdS_2/CQM$.
 \par

The object of interest in this paper is to compute HEE, HC and fidelity susceptibility  for a generic open string system using the $AdS_2/\mbox{open string}$ duality. The question of interest is how these quantities  evolve with time.

The structure of the paper is as follows. 
 In Sec. 2,  we fleetingly review the formal frameworks of  gravity on $AdS_2$. 
In Sec.  3,  we compute HEE of a string via RT formalism. In Sec. 4, we calculate HC of the string. In Sec. 5,  we investigate fidelity susceptibility holographically. In Sec. 6, we summarize our results.

\section{Gravity in two dimensions and $AdS_2/\mbox{open string}$ duality}
Let us start by two-dimensional dilaton theory with the following action,
\begin{eqnarray}
&&S=\frac{1}{2\kappa^2}\int d^2x \sqrt{-g}\Big(\phi R+V(\phi)\Big),\label{action}
\end{eqnarray}
where the potential is $V(\phi)=2\lambda^2 \phi$, $\lambda^2$ stands for cosmological constant $\Lambda$, and we redefine $\phi=e^{-2\varphi}$ where $\varphi$ is dilatonic field. A reason for breaking the conformakl symmetry is the existence of a non uniform profile for the dilaton scalar field  $\phi$ , i.e, when $\nabla_{\mu}\phi\neq0$ . It is remarkable to mention here that  the Birkhoff’s theorem holds generally , consequently we always can find static blackholes  \cite{Cadoni:2000ah}.  Let us consider 
an  asymptotically $AdS_2$ solution for dilaton action given in Eq. (\ref{action}). The trivial symmetry generators are a set of the  infinitesimal
diffeomorphisms. An equivalent description to these diffeomorphisms is given by the  \emph{pure gauge}. Note that these equivalence hiold only for regions near (almost existed) on the  boundary. It was demonstrated that these   symmetries  are governed by
a Virasoro algebra \cite{CM1}. 
The dilaton gravity action Eq. (\ref{action}) can be cast in a nonlinear conformal sigma model form. In Ref. \cite{M. Cavagli`a}
the duality between dilaton gravity action (\ref{action}) on $AdS_2$ and open strings was proved and a clear dualities exist between the two theories , one in the quantum system on boundary and the other as the geometry of a 
$AdS_2$ copy of the AdS spacetime.

The action given by (\ref{action}) basically has two solutions given by pure $AdS_2$ (ground state) and Schwarzschild-AdS (SAdS) with mass $M=m_{bh}$  which corresponds to the excited state in dual quantum theory. We write these solutions as following:
\begin{eqnarray}
&&\mbox{AdS},\ \  ds^2_{1}=\frac{1}{\lambda^2 x^2}(-dt^2+dx^2),\ \  \phi=\phi_0 \label{metric1}\\
&&\mbox{SAdS},\ \  ds^2_{2}=(\frac{a}{\sinh(a\lambda \sigma)})^2(-d\tau^2+d\sigma^2),\ \  a=\sqrt{\frac{2m_{bh}}{\lambda\phi_0}},\ \ \phi=\phi_0\lambda \sigma \label{metric2}.
\end{eqnarray}
The spacetime is the gravitational dual of an  open string with a conformal symmetry near its quantum critical point.

Note that the metric of pure AdS can be obtained as a limit of $\lim_{a\to 0} ds^2_{2}$, and the two metrics (\ref{metric2}), (\ref{metric1}) are related by the change of coordinates. However, this transformation can not cover all the spacetime manifold and is singular, so this transformation is formal and we won't use it in our investigation.

\section{Holographic Entanglement Entropy for String}
Using celebrated RT proposal \cite{6}-\cite{6a}, we need to calculate minimal area for two regions of entangled systems. We use time dependant formalism to calculate holographic entanglement entropy and holographic complexity \cite{Momeni:2016ira,Carmi:2017jqz}.
To compute entanglement entropy and complexity at the boundary using bulk we need to specify the boundary entangled region. The entangling region in the boundary is taken to be a string with width $L$ such
that
\begin{eqnarray}
&&\mbox{Pure AdS}=\mbox{A}_1=\{  t=t(x),\ \ -L\leq x\leq L\},\\&&
\mbox{SAdS}=\mbox{A}_2=\{  \tau=\tau(\sigma),\ \ -L\leq \sigma\leq L\},
\end{eqnarray}
where  $L$   is the extent of the subsystem in the time direction. Since the string has time translational invariance, one can describe the profile of the extremal surface by $t=t(x)$ for pure AdS and $\tau=\tau(\sigma)$ for blackhole. With this set up in place, we can now proceed to compute the RT area enclosed by the minimal surface extending from the boundary into the bulk.

\subsection{Pure AdS}
In this case, this is given  by 
\begin{eqnarray}
&&A_1(\gamma)=\int_{-L}^{L}\frac{dx}{|\lambda x|}\sqrt{1-\dot{t}^2},\label{func1}
\end{eqnarray}
where "dot" denotes derivative with respect to $x$. The minimization of this area functional determines the function $t(x)$ which reads,
\begin{eqnarray}
&&\dot{t}=\frac{|\lambda p x|}{\sqrt{1+(\lambda p x)^2}}\label{EL1}.
\end{eqnarray}
Boundary conditions are given by,
\begin{eqnarray}
&&t(x=0)=t^{*},\ \ \dot{t}(x=0)=0,\ \ t(x=L)=T-\epsilon\label{BC1}.
\end{eqnarray}
where $T$ denotes total time elapsed in system and $\epsilon$ is smallness parameter.  An Exact solution for (\ref{EL1}) is given as follows,
\begin{eqnarray}
&&t(x)=\frac{sgn(\lambda p x)}{\lambda p}\sqrt{1+(\lambda p x)^2}+C_1\label{tx}.
\end{eqnarray}
here $sgn(x)=\frac{x}{|x|}$. Using (\ref{BC1}) we obtain the ultimate form of solution for the extremal surfaces in $A_1$, as following,
\begin{eqnarray}
&&t(x)=\frac{sgn(\lambda p x)}{\lambda p}\sqrt{1+(\lambda p x)^2}+t^{*}\mp\frac{1}{|\lambda p|}.
\end{eqnarray}

Substituting the above expression for $t(x)$ in eq.(\ref{func1}) and putting a cut-off $p^{-1}$ for the $x$ integral, we have
 the following expression for HEE ,
\begin{eqnarray}
&&S_{HEE}^{AdS}=\frac{\ln(\lambda p L/2)}{\lambda}+\lim_{\sigma\to 0}\frac{\ln(\lambda p \sigma/2)}{\lambda}+
\mathcal{O}(p^2)\label{S1}.
\end{eqnarray} 
We will see when we substract (\ref{S1}) from the HEE for SAdS the divergence term will be cancelled.

\subsection{SAdS}
Now we calculate HEE for (\ref{metric2}). Area functional for (\ref{metric1}) is given by 
\begin{eqnarray}
&&A_2(\gamma)=\int_{-L}^{L}\frac{ad\sigma}{|\sinh(a\lambda\sigma)|}\sqrt{1-\dot{\tau}^2},\label{func2}
\end{eqnarray}
Euler-Lagrange Eq, is obtained as follows:
\begin{eqnarray}
&&\dot{\tau}=\frac{|\eta\sinh(a\lambda\sigma)|}{\sqrt{1+(\eta\sinh(a\lambda\sigma))^2}}\label{EL2}.
\end{eqnarray}
Here $\eta \ll 1$ is a conserved charge of system like $p$ in the last section.
Boundary conditions are given by,
\begin{eqnarray}
&&\tau(\sigma=0)=\tau^{*},\ \ \dot{\tau}(\sigma=0)=0,\ \ \tau(\sigma=L)=T^{*}-\epsilon\label{BC2}.
\end{eqnarray}
$T^{*}$ denotes the total time elapsed in system and $\epsilon$ is smallness parameter. Exact solution for (\ref{EL2}) is given by follows,
\begin{eqnarray}
&&\tau(\sigma)=\tau^{*}-\frac{\eta}{a\lambda|\eta|}\ln(1+\eta)+\frac{\eta\sinh(a\lambda\sigma)}{a\lambda|\eta\sinh(a\lambda\sigma)|}\\&&
\times\Big(\eta\cosh(a\lambda\sigma)
+\sqrt{1+\eta^2\sinh^2(a\lambda\sigma)
}\label{taux}
\Big).
\end{eqnarray}
 and $\eta\to 0$. Expanding $\tau(\sigma)$ in a Laurent series in $\eta$, we find the following expression for HEE,
\begin{eqnarray}
&&S_{HEE}^{SAdS}=\tanh^{-1} \left(\frac{1}{2}\,{\frac {1+ \left( \cosh \left( a\lambda\,L
 \right)  \right) ^{2}}{\cosh \left( a\lambda\,L \right) }} \right) 
+\lim_{\eta\to0}\frac{4}{a\lambda\eta}+\mathcal{O}(\eta^2)
\label{S2}.
\end{eqnarray} 
So, the total, finite HEE is given by Eq. (\ref{S2})-(\ref{S1}), 
\begin{eqnarray}
&&S_{HEE}^{Net}=S_{HEE}^{SAdS}-S_{HEE}^{AdS}=\frac{1}{4G}\Big(\tanh^{-1} \left(\frac{1}{2}\,{\frac {1+ \left( \cosh \left( a\lambda\,L
 \right)  \right) ^{2}}{\cosh \left( a\lambda\,L \right) }} \right)\Big)\\&&\nonumber -\frac{\ln(\lambda p L/2)}{4 G\lambda}
+\frac{1}{4G}\lim_{\eta\to0}\Big(\frac{4}{a\lambda\eta}-\frac{\ln(\frac{\lambda p \eta}{2})}{\lambda}\Big)
\end{eqnarray}
 The leading divergent term of the HEE of an extremal co-dimension one hypersurface in the $AdS_2$ geometry is,
 \begin{eqnarray}
 &&S^{div}_{HEE}\sim \frac{1}{G\lambda}\Big(\frac{1}{a\eta}-\ln(\frac{\lambda^2 Lp^2 \eta}{4})\Big)
 \end{eqnarray}
where $\eta\to 0,p\to\infty$. If we define $b\equiv \eta p<\infty$, the divergent term is rewritten as follows:
\begin{eqnarray}
&&S^{div}_{HEE}\sim \frac{1}{G\lambda}\Big(\frac{1}{a\eta}+\ln(\eta)+\mbox{finite term}\Big).
\end{eqnarray}
We have found that HEE is scaled as follows:
\begin{eqnarray}
&&S^{AdS_2,div}_{HEE}\sim \ln(\eta)\\&&
S^{SAdS_2,div}_{HEE}\sim \frac{1}{\eta}
\end{eqnarray}

\section{Computation of Holographic complexity}
In this section we calculate holographic complexity for metrics (\ref{metric1},\ref{metric2}) using  the RT volume enclosed by the minimal surface extending from the boundary into the bulk \cite{Momeni:2016ira}.
\subsection{Pure AdS}
For metric (\ref{metric1}) the RT volume enclosed by the minimal surface is given by the following integral,
\begin{eqnarray}
&&V_1=\int_{0}^{L}\frac{dx}{\lambda^2 x^2}\Big[t(x)-t^{*}\Big]
\end{eqnarray}
Using solution in Eq. (\ref{tx}) and by change of variables: $y=\lambda p x$ and $y_0=\lambda p L$, we obtain:
\begin{eqnarray}\label{V1}
&&V_1=\frac{1}{\lambda^2}\Big(\sinh^{-1}y_0+y_0
\sqrt{1+y_0^2}-\frac{(1+y_0^2)^{3/2}}{y_0}+\frac{1}{y_0|\lambda p|}
\Big)\\&&\nonumber
-\frac{1}{\lambda^2}\lim_{\epsilon\to 0}(\frac{1}{|\lambda p|}-1)\frac{1}{\epsilon}
\end{eqnarray}
Note that we used taylor's series for integral near $y=\epsilon\to 0$.  The divergent term of HC is rewritten as following:
\begin{eqnarray}
&&\mathcal{C}^{AdS_2}\sim \frac{1}{\epsilon}.
\end{eqnarray}

\subsection{SAdS}
For SAdS, using metric (\ref{metric2}) and by solution given in Eq. (\ref{taux}) we obtain:
\begin{eqnarray}
&&V_2=\frac{\Phi(z_0,\epsilon)}{\lambda^2}+\frac{1}{\lambda^2}\frac{\eta\ln(1+|\eta|)}{|\eta|}\frac{\cosh z_0}{\sinh z_0}-\frac{1}{\lambda^2}\lim_{\epsilon\to 0}\frac{\eta\ln(1+|\eta|)}{|\eta|}\frac{1}{\epsilon}\label{V2}
\end{eqnarray}
where $z_0=\lambda a L$ and we define an auxiliary function 
\begin{eqnarray}
&&\Phi(z_0,\epsilon)=-\frac{\ln(1+\frac{\eta}{|\eta|})}{z_0}+\frac{\ln(1+\frac{\eta}{|\eta|})}{\epsilon}+\mathcal{O}(z_0)
\end{eqnarray}
where the subleading terms start at order $\mathcal{O}(\frac{1}{\epsilon})$. 
Finally if we substract (\ref{V1}) from (\ref{V2}) we obtain a finite value for holographic complexity,
\begin{eqnarray}
&&\Delta\mathcal{V}=-\frac{\ln(1+\frac{\eta}{|\eta|})}{\lambda^3 a L}+\frac{1}{\lambda^2}\frac{\eta\ln(1+|\eta|)}{|\eta|}\frac{\cosh (\lambda a L)}{\sinh (\lambda a L)}\\&&\nonumber-\frac{1}{\lambda^2}\Big(\sinh^{-1}\lambda p L+\lambda p L
\sqrt{1+(\lambda p L)^2}-\frac{(1+(\lambda p L)^2)^{3/2}}{\lambda p L}+\frac{1}{\lambda p L|\lambda p|}
\Big)
\end{eqnarray}

The divergent term of HC is rewritten as follows:
\begin{eqnarray}
&&\mathcal{C}^{SAdS_2}\sim \frac{1}{\epsilon}.
\end{eqnarray}
where the divergenvt parts is removed safely if diffrent conserved charges $(p,\eta)$ satisfies the following (resembeling an IR-UV cut off relation),
\begin{eqnarray}\label{UV-IR}
&&\Big(1+\frac{\eta}{|\eta|}\Big)
\Big(1+|\eta|\Big)^{-\frac{\eta}{|\eta|}}=\exp(1-\frac{1}{|\lambda p|}).
\end{eqnarray}
To further justify this UV-IR relation Eq. (\ref{UV-IR}), it is constructive to extend our discussions for the case where the boundary has  a massles scalar field $\psi$ and a massive scalar field  $\Psi$ with mass $M$. Obviously if we keep the energy level $p\ll M$, we won't see the massive field, $p$ is the IR cutoff. If we now take the energy to be $\eta\gg M$
we will see both fields $(\Psi,\psi)$, but now almost all of their energy will be momentum and their mass will be negligible. They will be behaving as two massless fields. This is an example of hierarchical UV/IR mixing \cite{Lust:2017wrl},

\section{Holographic Fidelity susceptibility}
In previous section, we proved that quantum complexity of an open string theory can also be obtained holographically, as the holographic complexity is dual to a volume in $AdS_2$ space time. Furthermore, it has been demonstrated that the holographic complexity of a field theory can be proportional to the fidelity susceptibility of the boundary field theory. Consequently, the fidelity susceptibility of an open string theory can be holographically calculated using a maximal volume $V_{max}$ in the $AdS_2$ which ends on  a covariant zero mean curvature slicing of the time-dependent bulk geometry \cite{MIyaji:2015mia}. We can use this maximal volume in the AdS geometry to define the holographic complexity in such a geometry as
\begin{eqnarray}
&&F=\frac{V_{max}}{8\pi R G}.
\end{eqnarray}
Now the $AdS_2/\mbox{open string}$ holography is a limiting case of  usual holography , and it is known that there are divergence terms associated with such volumes. Thus, we need to regularize this volume, before we can define the fidelity susceptibility for $AdS_2$ geometries. This will be done by subtracting the background AdS geometry $V_{max}^{AdS}$ from the deformed SAdS geometry $V_{max}^{SAdS}$. So, we can define a regularized maximal volume
\begin{eqnarray}
&&\Delta \mathcal{V}_{max}=V_{max}^{SAdS}-V_{max}^{AdS}.
\end{eqnarray}
Now using this regularized maximal volume in the $AdS_2$ geometry, we can define the regularized holographic fidelity susceptibility of an open string boundary theory as
\begin{eqnarray}
&&\Xi_F=F_{SAdS}-F_{AdS}=\frac{\Delta\mathcal{V}_{max}}{8\pi R G}\label{xi}
\end{eqnarray}
where $R$ is the radius of the curvature of this $AdS_2$ geometry. This regularized holographic fidelity susceptibility is equal to fidelity susceptibility of the boundary open string theory, and so the fidelity susceptibility of the open string field theory can be holographically estimated from holographic complexity. It may be noted that recently in Ref. \cite{Alishahiha:2017cuk}, the author demonstrated that the holographic fidelity susceptibility for two quantum states  is given by the difference of the volume of an extremal surface in the fully back reacted dual  geometry.
Now we can use the Eq. (\ref{xi})  to evaluate fidelity susceptibility as
\begin{eqnarray}
&&\Xi_F=\frac{1}{8\pi R G}\Big(\frac{T-t^{*}}{L}-\frac{a(T^*-\tau^*)}{\lambda}\frac{\cosh(a\lambda L)}{\sinh(a\lambda L)}\Big).
\end{eqnarray}
where the divergene parts are removed .\par
To conclude one may propose that the renormalized fidelity susceptibility of the maximal volume due to the effect of a time interval  cancelation provide a holographic Kepler's second law  for fidelity susceptibility in the dual field theory
\begin{eqnarray}
&& T^{*}-\tau^*=T-t^*.
\end{eqnarray}
\emph{Conjecture}:{\it An imaginary line connecting two points in an open string in dual $AdS_2$, sweeps out an equal fidelity susceptibility  of quantum system in equal amounts of time.}

\section{Summary}
In this letter,  we calculate  holographic
quantum complexity and the holographic entanglement entropy for string holographically. 
We propose that a hierarchical UV/IR mixing exists in $AdS_2/\mbox{open strings}$. It can 
be considered  as a  holographic version of Kepler's  second law argues that an imaginary line connecting two points in an open string in dual $AdS_2$, sweeps out an equal fidelity susceptibility  of quantum system in equal amounts of time. Furthermore, in analogy with  the usual   Kepler's  second law, the regions 
analysed were assumed to  exist as connected points in string. We argued that such a conjuncture should hold in general, as it is based on 
the AdS version of the Kepler's  second law.

\section*{Acknowledgments}

The work of KB is partially supported by the JSPS KAKENHI Grant Number JP 25800136 and Competitive Research Funds for Fukushima University Faculty (17RI017).

\end{document}